\documentclass[aps,prb,superscriptaddress,showpacs, preprint
]{revtex4}
\usepackage{graphicx}
\usepackage{epstopdf}
\usepackage{amssymb,amsmath,amsfonts}

\providecommand{\abs}[1]{\lvert#1\rvert}
\newcommand{\coso} {{\mbox{Cu${}_2$OSeO${}_3$}}}
\newcommand{\ca} {{\mbox{Cu${}^{\mathsf{I}}$}}}
\newcommand{\cb} {{\mbox{Cu${}^{\mathsf{II}}$}}}
\begin{document}
\title{Two-step transition in a magnetoelectric ferrimagnet $\coso$}
\author{I. \v Zivkovi\'c}
\email{zivkovic@ifs.hr}
\affiliation{Institute of Physics, P.O.B.304, HR-10 000, Zagreb, Croatia}
\author{D. Paji\'c}
\affiliation{Department of Physics, Faculty of Science, Bijeni\v cka c.32, HR-10 000 Zagreb, Croatia}
\author{T. Ivek}
\affiliation{Institute of Physics, P.O.B.304, HR-10 000, Zagreb, Croatia}
\author{H. Berger}
\affiliation{Institute of Physics of Complex Matter, EPFL, CH-1015 Lausanne, Switzerland}

\date{\today}

\begin{abstract}
We report a detailed single crystal investigation of a magnetoelectric ferrimagnet $\coso$ using dc magnetization and ac susceptibility along the three principal directions [100], [110] and [111]. We have observed that in small magnetic fields two magnetic transitions occur, one at $T_c = 57$ K and the second one at $T_N = 58$ K. At $T_c$ the non-linear susceptibility reveals the emergence of the ferromagnetic component and below $T_c$ the magnetization measurements show the splitting between field-cooled and zero-field-cooled regimes. Above 1000 Oe the magnetization saturates and the system is in a single domain state. The temperature dependence of the saturation below $T_c$ can be well described by $\mu(T) = \mu(0)[1-(T/T_c)^2]^\beta$, with $\mu(0) = 0.56 \mu_B$/Cu, corresponding to the 3-up-1-down configuration. The dielectric constant measured on a thin single crystal shows a systematic deviation below the transition, indicating an intrinsic magnetoelectric effect.
\end{abstract}


\maketitle

%
%
%
%

\section{Introduction}
\label{Introduction}

Although first investigations of magnetoelectrics have been performed in 1960s~\cite{Rado1961}, a revitalization of the interest in this type of materials came with the discovery of a colossal magnetoelectric effect in manganites $R$MnO$_3$~\cite{Kimura2003} and $R$Mn$_2$O$_5$~\cite{Hur2004}. Since then numerous materials have been investigated with different types of coupling between the electric and magnetic order parameters~\cite{Khomskii2009}. The prospect of using these materials in new types of memory elements where the magnetic order could be easily manipulated with the applied voltage is stimulating further investigation of new materials.

Recently, it has been shown that $\coso$, a ferrimagnet below $\approx 60$ K, shows a magnetoelectric effect below the magnetic transition~\cite{Bos2008}. This has been revealed through the change of the dielectric constant across the magnetic transition~\cite{Bos2008} and later confirmed by infrared measurements~\cite{Miller2010,Gnezdilov2010}. $\coso$ crystallizes in a cubic space-group $P2_13$ and a high resolution structural investigation revealed that it remains metrically cubic down to 10~K~\cite{Bos2008}. There are two crystallographically inequivalent copper sites, $\ca$ within a trigonal bipyramidal $\ca$O$_5$ unit and $\cb$ within a square pyramidal $\cb$O$_5$ unit. Both copper ions are in a 2+ oxidation state, resulting in one unpaired electron per site, giving rise to spin S = 1/2 on each magnetic ion. The observed ferrimagnetic structure is a result of a $\ca : \cb = 1 : 3$ ratio, with all the moments pointing along the space diagonal~\cite{Bos2008}. From the neutron scattering experiments it has been found that the coupling between the nearest-neighbor (NN) moments is antiferromagnetic for $\ca$ -- $\cb$ and ferromagnetic for $\cb$ -- $\cb$~\cite{Bos2008}, which has been later confirmed by the NMR study~\cite{Belesi2010}.

The exact nature of the magnetoelectric coupling in $\coso$ is still unclear and deeper insight is needed in the basic magnetic properties of the system. It is especially important to elucidate the details of the magnetic transition where the dielectric measurements on powdered sample showed the critical behavior~\cite{Bos2008}. In this report we present detailed dc magnetization and ac susceptibility measurements along three principal directions in the cubic system ([100], [110] and [111]). Our results show that in zero field the domain structure becomes important, giving rise to a complicated M -- H dependence as the field is applied. A rather small field of 1000 Oe is enough to drive the system to saturation, where the value of the magnetization of the plateau follows the same temperature dependence found for the internal field by $\mu$SR ~\cite{Maisuradze2011}. However, in zero field the susceptibility above the transition does not show the critical behavior. Also, our dielectric measurements did not reveal a strong kink at the transition, as observed for the powdered sample~\cite{Bos2008}. On the other hand, below the transition the dielectric constant deviates from the high temperature behavior, supporting the conclusion about the magnetoelectric effect in $\coso$.

%
%
%
%

\section{Experimental details}
\label{Details}

Single crystals have been prepared by the chemical vapor phase method. Details of the preparation can be found in previous reports~\cite{Miller2010,Belesi2010}. The orientation of single crystals has been performed using the x-ray Laue camera after which the samples were cut in shapes with dimensions 1x1x4 mm$^3$ with the longest side along [100], [110] and [111].

Magnetization measurements were performed on a commercial Quantum Design MPMS5 magnetometer with a temperature range  2 K -- 800 K and magnetic fields up to 5 T. The ac susceptibility was measured using a commercial CryoBIND system with a temperature range 4 K -- 400 K and a frequency range 10 mHz -- 10 kHz. The dielectric response has been measured using the home-made setup with the HP 4284A LCR meter in the frequency range 20 Hz -- 1 MHz, with excitation voltages of 50 mV and 1 V. The sample used has a plan-parallel shape with a cross section of 2.8 mm$^2$ and a thickness of 0.4 mm. The contacts have been prepared using the DuPont 4929N silver paint.

%
%
%
%

\section{Results}
\label{Results}

\begin{figure*}
\includegraphics[width=0.9\textwidth]{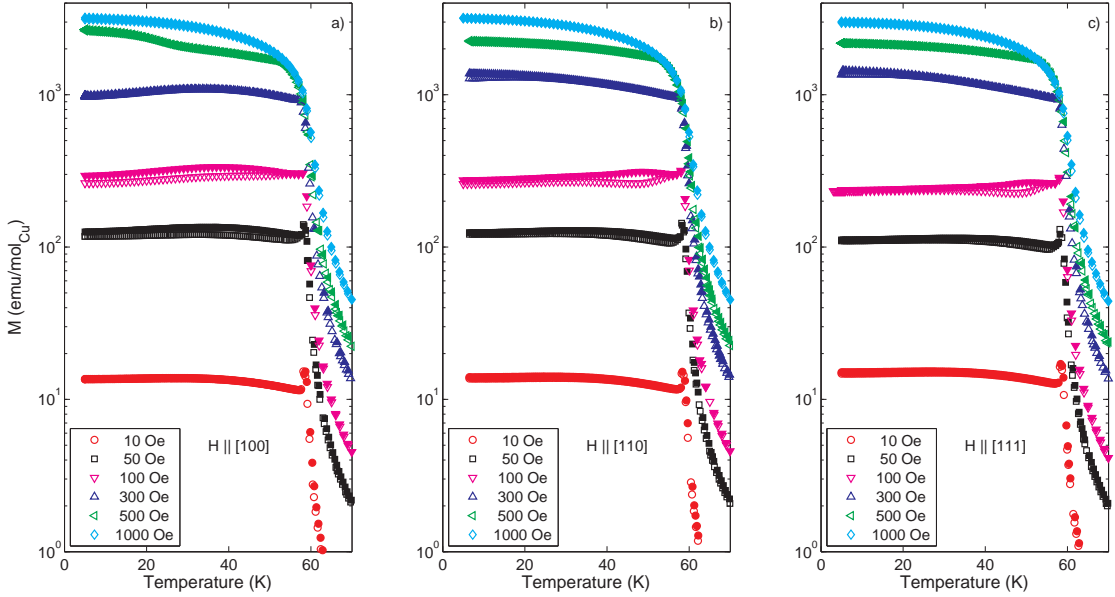}
\caption{(Color online) Temperature dependence of the magnetization for a) [100], b) [110] and c) [111] directions. Upper branches (full symbols) correspond to the FC regime while the lower branches (empty symbols) correspond to the ZFC regime.}
\label{fig-MT}
\end{figure*}

In Fig.~\ref{fig-MT} we present a detailed investigation of the temperature dependence of the magnetization ($M$) for various applied fields below the transition for all three principal directions. Two sets of measurements are presented for each field, field-cooled (FC) and zero-field-cooled (ZFC) curves. The procedure for ZFC measurements includes cooling the sample in zero field, applying the field and measuring while warming up. FC measurements have been performed while cooling the sample with the field applied.

A noticeable feature is the development of the difference between ZFC and FC curves. For a small field ($H \approx 10$ Oe) no difference is observed. As the field is increased ($H \approx 100$ Oe), the splitting starts below the transition, goes through a maximum around 40 K -- 50 K and then decreases. For even larger fields ($H \approx 500$ Oe) ZFC and FC curves overlap down to 20 K -- 30 K and then gradually split below 20 K. Finally, for fields $H \gtrsim 1000$ Oe again no splitting occurs. These observations are in accordance with the results from the powder investigation~\cite{Bos2008} where an opening of a small hysteresis loop has been reported in the range 300 Oe -- 500 Oe.

Out of the three principal directions, [100] shows a somewhat different behavior compared to [110] and [111] (which are qualitatively very similar) regarding the development of the FC -- ZFC splitting. For $H = 10$ Oe and 1000 Oe fields all three directions show almost identical behavior. However, already for 50 Oe the measured FC -- ZFC difference is larger for the [100] direction. With 100 Oe it persists down to lowest temperatures with $H || [100]$ while for the other two directions it decreases towards zero (more so for $H || [111]$). For $H = 300$ Oe $M^{[100]}$ reveals a broad maximum around 40 K where $M^{[110]}$ and $M^{[111]}$ continue to increase down to 5 K. Finally, with $H = 500$ Oe, there is a change in the slope in $M^{[100]}$ around 25 K. All this points to a very complicated mechanism of magnetization processes which is very sensitive to small fields and the orientation of the sample.

\begin{figure}
\includegraphics[width=0.4\textwidth]{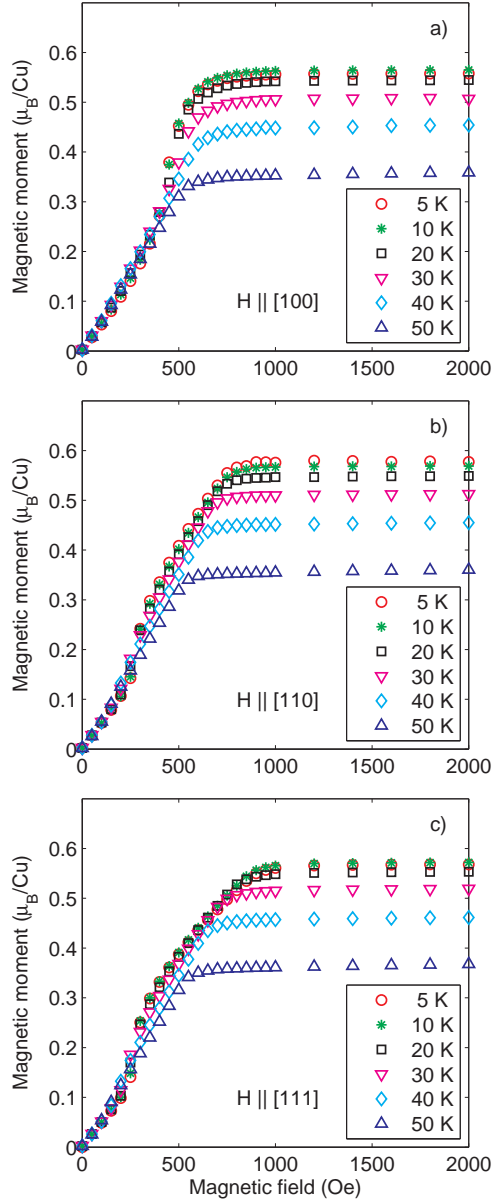}
\caption{(Color online) Average magnetic moment on the cooper ion vs. magnetic field for a) [100], b) [110] and c) [111] directions.}
\label{fig-MH}
\end{figure}

To gain insight into this complicated field behavior, we have measured the field dependence of the magnetization. Each curve in Fig.~\ref{fig-MH} has been recorded after the sample has been cooled down from above the transition with the applied magnetic field of $H = 2000$ Oe (FC regime).

All the curves exhibit a quasi-linear regime below 600 -- 1000 Oe (depending on temperature) above which the saturation plateau is reached. Additionally, as the temperature is lowered below 50 K a metamagnetic transition occurs around 250 Oe for the [110] and [111] directions. [100] direction is again somewhat different qualitatively and quantitatively, with a characteristic field above 400 Oe.

The saturation values at all temperatures are identical for all three directions. This means that the sample becomes a single domain for relatively small applied fields ($\approx 1000$ Oe at 5 K) which suggests that $\coso$ exhibits a Heisenberg anisotropy to a very good approximation.

\begin{figure}
\includegraphics[width=0.45\textwidth]{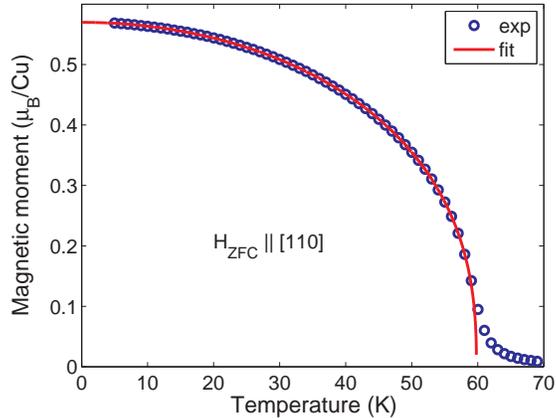}
\caption{(Color online) Temperature dependence of the saturated magnetization in 1000 Oe for the [110] direction. The line is a fit to Eq.~(\ref{eq-pL})}
\label{fig-critical}
\end{figure}

For that reason it is interesting to look more closely into the temperature dependence of the average magnetic moment per Cu ion $\mu (T)$ when the sample is saturated. In Fig.~\ref{fig-critical} we plot the results for $H \parallel [110]$. The solid line represents the fit to a power law
\begin{equation}
\mu(T) = \mu(0)\left[ 1 - \left( \frac{T}{T_c} \right)^\alpha \right]^\beta
\label{eq-pL}
\end{equation}
which has also been used to describe the internal field behavior in the $\mu$SR study~\cite{Maisuradze2011}. If the parameter $\alpha$ is varied in the fitting procedure, it yields the value $\alpha = 1.95$, very close to 2.00 obtained in the $\mu$SR study~\cite{Maisuradze2011}. $\alpha = 2$ is supported by the general observation that the variation of the order parameter close to $T = 0$ K depends only on the dimensionality and the value of the spin quantum number~\cite{Koebler1999}. Thus, we have fixed $\alpha = 2$ and allowed other parameters to be determined by the fit. The fitting has been performed on the total of six curves (three directions, FC and ZFC conditions). Averaging all the results gives $\mu(0) = 0.559(7) \mu_B$, $T_c^{critical} = 59.85(5)$ K and $\beta = 0.393(4)$ for the saturation value, critical temperature and critical exponent, respectively.

The value of the critical exponent $\beta$ agrees with the value deduced by $\mu$SR~\cite{Maisuradze2011}. It is very close to the theoretical value for a 3D Heisenberg system ($\beta^{Heis. 3D} = 0.365$) and it is in excellent agreement with the values obtained for various shapes of the polycrystalline Ni samples~\cite{Seeger1995}.

$T_c^{critical} = 59.85$ K is somewhat higher than deduced from the $\mu$SR study (57 K). The crucial difference is that in our case the sample experienced an external magnetic field of $H = 1000$ Oe while the $\mu$SR study has been performed in a zero-field condition. In a recent study of the field dependence of the transition~\cite{Huang2011} it has been revealed that there is a rather strong positive coefficient where $T_c$ increases by 10 K in $H = 7$ T. However, their conclusion was that the transition remains immobile up to 1 T. On the other hand, their method involved the derivative of the magnetization which is not sensitive enough to reliably determine such a small change.

The saturation value at low temperatures $\mu(0) = 0.559(7) \mu_B$ corresponds to the 3-up-1-down configuration where 3-up is associated with $\cb$ ions and 1-down with a $\ca$ ion. The remarkable fact is that Eq.~\ref{eq-pL} describes the data in the whole temperature range, with the critical behavior for $T \rightarrow 0$ and $T \rightarrow T_c$, which is rarely found even in much simpler systems.

\begin{figure}
\includegraphics[width=0.45\textwidth]{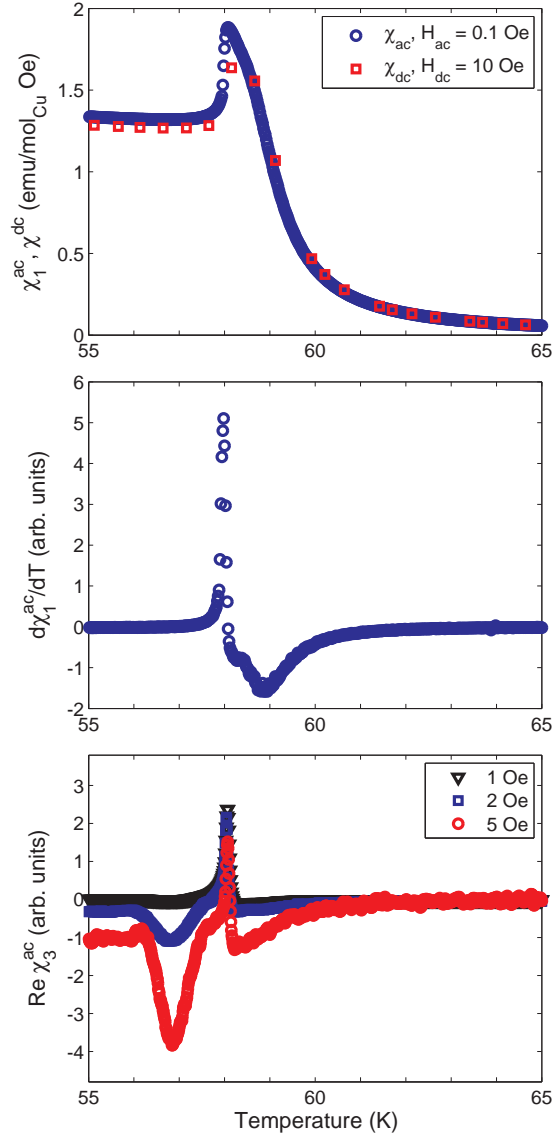}
\caption{(Color online) Temperature dependence of the susceptibility around the transition for the [111] direction. Top panel: linear susceptibility (ac and dc). Middle panel: a derivative of the linear ac susceptibility. Bottom panel: the 3rd harmonic of the ac susceptibility.}
\label{fig-acdc}
\end{figure}

Now it is interesting to look more closely at how the system transforms from the paramagnetic to the ferrimagnetic state. In Fig.~\ref{fig-acdc} we show the expanded view around 60 K, covering the transition to the ordered state. In the upper panel we present the comparison between the ac and dc results in small fields, measured with 0.1 Oe and 10 Oe, respectively. The ac susceptibility $\chi_1^{ac}$ has been normalized so that it matches the dc susceptibility $\chi^{dc}$ in the paramagnetic regime above 70 K but one can notice that they overlap even lower, down to $\approx 62$ K. No frequency dependence of the ac susceptibility has been observed, confirming the previous finding~\cite{Huang2011}. The non-linear response, measured with a 3rd harmonic of the ac susceptibility $\chi_3^{ac}$, is shown in the lower panel.

Since the ac susceptibility is measured in the sweep mode, it is easier to notice the subtle features often missed in step-wise curves, usually associated with the extraction method. One can notice that the divergent-like behavior of the susceptibility, expected from a ferrimagnet, exists down to 59 K but then it turns into an S-shape curve and the criticality is avoided. This is even more evident when the derivative of the susceptibility is calculated (see the middle panel of Fig.~\ref{fig-acdc}). The slope of the curve gets more and more negative until it reaches the maximum negative value just below 59 K, marking the presence of an inflection point. In comparison, the dc susceptibility data show the ambiguity even for the determination of the transition temperature due to a relatively large step between points (0.5 K in our case, cf. Ref.~\onlinecite{Miller2010} with a 1 K step). At $T_N = 58$ K a sharp drop is observed, marking the transition to the ordered state. It is instructive to mention that the same behavior of the ac susceptibility is observed for all the investigated directions.

In the lower panel of Fig.~\ref{fig-acdc} we present the results of the non-linear susceptibility measurements. This technique is recognized as a powerful tool in the description of magnetic transitions, first from the theoretical point of view~\cite{Fujiki1981} and then with an implementation for ferromagnets~\cite{Bitoh1993,Nair2003}, antiferromagnets~\cite{Narita1996}, spin-glasses and superparamagnets~\cite{Bitoh1993a}. One can categorize the system by its dependence on the field amplitude and the observation of divergences on both sides of the transition.

The shape of the non-linear response around $T_N = 58$ K strongly suggests an antiferromagnetic nature of the ordered state~\cite{Fujiki1981,Narita1996}: it is diverging below $T_N$ but non-diverging above $T_N$. A surprising discovery is the presence of a second peak in the non-linear part of the susceptibility $\chi_3^{ac}$, centered at $T_c = 57$ K. This is accompanied by a broad minimum in $\chi_1^{ac}$ after which a slow increase of $\chi_1^{ac}$ is observed as the temperature is lowered. The peak magnitude increases with an increase of the amplitude of the driving field, suggesting the emergence of a ferromagnetic component in the system, in line with the observation of the appearance of the internal field from the $\mu$SR study~\cite{Maisuradze2011}.

\begin{figure}[h!]
\includegraphics[width=0.45\textwidth]{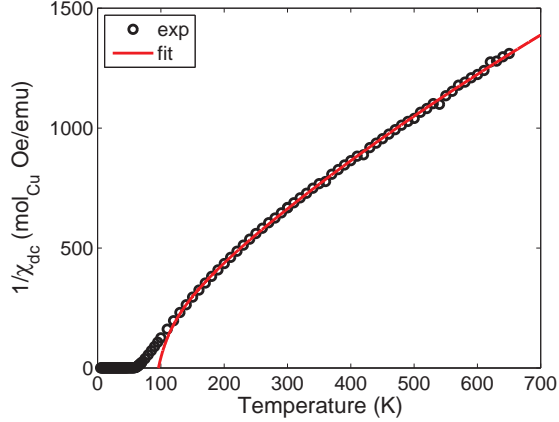}
\caption{(Color online) Temperature dependence of the inverse dc susceptibility. The line is a fit to Eq.~(\ref{eq-MF}).}
\label{fig-MF}
\end{figure}

Above the transition the system is in a paramagnetic regime. It has been suggested~\cite{Belesi2010} that it can be described within a mean-field approach, taking into account one ferromagnetic ($\cb$ -- $\cb$) and one antiferromagnetic ($\ca$ -- $\cb$) interaction. The susceptibility for $T > T_C$ is given by the expression
\begin{equation}
\chi = \frac{g^2S(S + 1)\mu_B^2}{3k_B(T - T_C)} \, \, \frac{T - \frac{1}{4}(\abs{J_{FM}} + 3J_{AFM})}{T + T_C - \abs{J_{FM}}} + \chi_0
\label{eq-MF}
\end{equation}
where $J_{FM} \approx -52$ K, $J_{AFM} \approx 75$ K, $\chi_0 = 1.3 \cdot 10^{-4}$ emu/mol Oe and the mean-field critical temperature is given by $T_C = (\abs{J_{FM}} + \sqrt{J_{FM}^2 + 3J_{AFM}^2})/2 \approx 90$ K. As presented in Fig.~\ref{fig-MF}, this model describes well the experimental data above 130 K.

\begin{figure}
\includegraphics[width=0.45\textwidth]{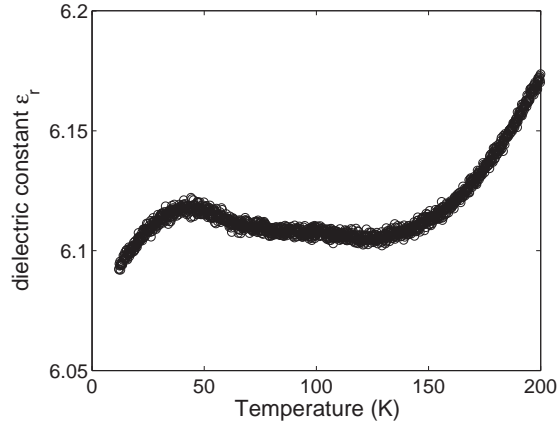}
\caption{The temperature dependence of the dielectric constant of $\coso$ measured at 10 kHz and 1 V.}
\label{fig-dielectric}
\end{figure}

The initial report by Bos and co-workers~\cite{Bos2008} suggested that the magnetoelectric effect emerges coincidently with the magnetic transition occurring at 60 K. Given the fact that we have revealed the existence of two magnetic transitions, it is crucial to try to connect one of them to the dielectric properties of the system. To this end, we have performed a similar investigation of the dielectric properties. At all temperatures we observe a purely real dielectric response with no significant dependence on frequency in our measurement range. In Fig.~\ref{fig-dielectric} we present the results of the measurement performed with 10 kHz and 1 V excitation. Very similar results have been obtained at lower frequencies and a lower excitation value of 50 mV, except for an increase in the noise level.

Compared with the results on the powder~\cite{Bos2008}, our data show somewhat different behavior. $\epsilon_r$ levels off below 130 K and exhibits a maximum around 45 K. We are unable to identify the presence of a kink around the magnetic transition from which the critical behavior of $\epsilon_r$ has been extracted~\cite{Bos2008}. However, below 45 K $\epsilon_r$ starts to decrease, very similar to the previous report~\cite{Bos2008}. Also, the values of the dielectric constant are different, we obtain $\epsilon_r \approx 6.1$ while for the powdered sample it has been obtained that $\epsilon_r \approx 14.3$.

The discrepancies between our single crystal investigation and the results of Bos and co-workers on the powdered sample can have several explanations. First, the measurements on the powdered sample are prone to a parasitic inter-grain capacitance, the so-called Maxwell-Wagner effect. Second, the retentive properties of the silver-paint applied to the pellet and to the smooth surface of the single crystal are very different. One has to bear in mind that, at least in our case, the total change in capacitance below 150 K corresponds to only 1 femtoFarrad. Thus, even a minute change in the contact properties or even in cables could affect the measurement. Although our results were reproducible in several cooling cycles, we cannot exclude the possibility of a reversible change in contacts and its extrinsic contribution. On the other hand, the observation of a maximum around 45 K and a subsequent drop of $\epsilon_r$ at lower temperatures is seen in both the powdered and single crystal form, indicating an intrinsic effect.

%
%
%
%

\section{Discussion}
\label{Discussion}

Previous reports suggested a single magnetic transition from a paramagnetic to a ferrimagnetic state occurring around 60 K which was accompanied by a change in the dielectric constant, revealing a magnetoelectric effect~\cite{Bos2008}. However, the ac susceptibility data shown in Fig.~\ref{fig-acdc} show that the transition to the ordered state of $\coso$ is not a simple one and requires a deeper insight.

We have found the signatures of two transitions, one at $T_c = 57$ K and one at $T_N = 58$ K. At $T_c$ the non-linear susceptibility revealed that a ferromagnetic component emerges in the system which is in agreement with previous neutron~\cite{Bos2008} and $\mu$SR~\cite{Maisuradze2011} studies. We can consider that for $T < T_c$ the system is in a ferrimagnetic 3-up-1-down state with moments pointing along the space diagonal~\cite{Bos2008}. The cubic symmetry is maintained down to the lowest temperatures~\cite{Bos2008,Belesi2010} which means that there are 4 equivalent space diagonals and each diagonal can have 2 directions of the composite moment. The minimization of the free energy will result in the creation of domains with 8 preferential directions. The rise of the linear component of the susceptibility below $T_c$ for small fields can then be understood as a consequence of the dynamics of domain walls.

Let us now discuss the possible origin of the transition at $T_N$. It is characterized by a sharp drop in the linear component of the susceptibility, observed for all the investigated directions ([100], [110] and [111]). The non-linear component of the susceptibility displays a divergent-like branch below $T_N$ and a non-divergent branch above $T_N$, which are characteristic features for a transition to the antiferromagnetic state~\cite{Fujiki1981,Narita1996}. However, for a simple, canonical easy-axis antiferromagnet~\cite{Ashcroft1976} one would expect for linear components $\chi_1^{[100]}$ and $\chi^{[110]}$, although not perpendicular to [111], to show a noticeable difference compared to $\chi_1^{[111]}$.

The mean-field approach, suggested by Belesi and coworkers~\cite{Belesi2010}, takes into account only two types of interactions, $\ca$ -- $\cb$ (AFM) and $\cb$ -- $\cb$ (FM). However, a careful inspection of the crystal structure reveals that there are several Cu -- O -- Cu bridges that can influence the magnetic behavior of the system. In Table~\ref{tb-CuO} we list the associated angles and distances, grouped to form NN interactions. According to the Goodenough-Kanamori-Anderson rule, the exchange interaction $J$ changes from AFM to FM as the Cu -- O -- Cu angle approaches $90^0$. The exact angle where $J$ changes sign lies in the range $95^0$ -- $99^0$ and is also dependent on the influence of the neighboring orbitals~\cite{Mizuno1998}.

\begin{table}
\caption{Cu -- O -- Cu bridges in $\coso$. The bridges between the horizontal lines form a single magnetic interaction.}
\label{tb-CuO}
\begin{tabular}{|c|c|c|}
\hline
type &  angle (deg) & distance (\AA) \\
\hline
$\ca$ -- O(2) -- $\cb$ & 117.9 & 1.909/2.038 \\ 
\hline
$\ca$ -- O(1) -- $\cb$ & 107.8 & 1.925/1.854 \\ 
$\ca$ -- O(3) -- $\cb$ & 94.15 & 2.082/2.089 \\ 
\hline
$\cb$ -- O(1) -- $\cb$ & 111.05 & 1.854 \\ 
\hline
$\cb$ -- O(2) -- $\cb$ & 99.92 & 2.038 \\ 
$\cb$ -- O(4) -- $\cb$ & 94.77 & 1.896 \\ 
\hline
\end{tabular}
\end{table}
\begin{figure}
\includegraphics[width=0.45\textwidth]{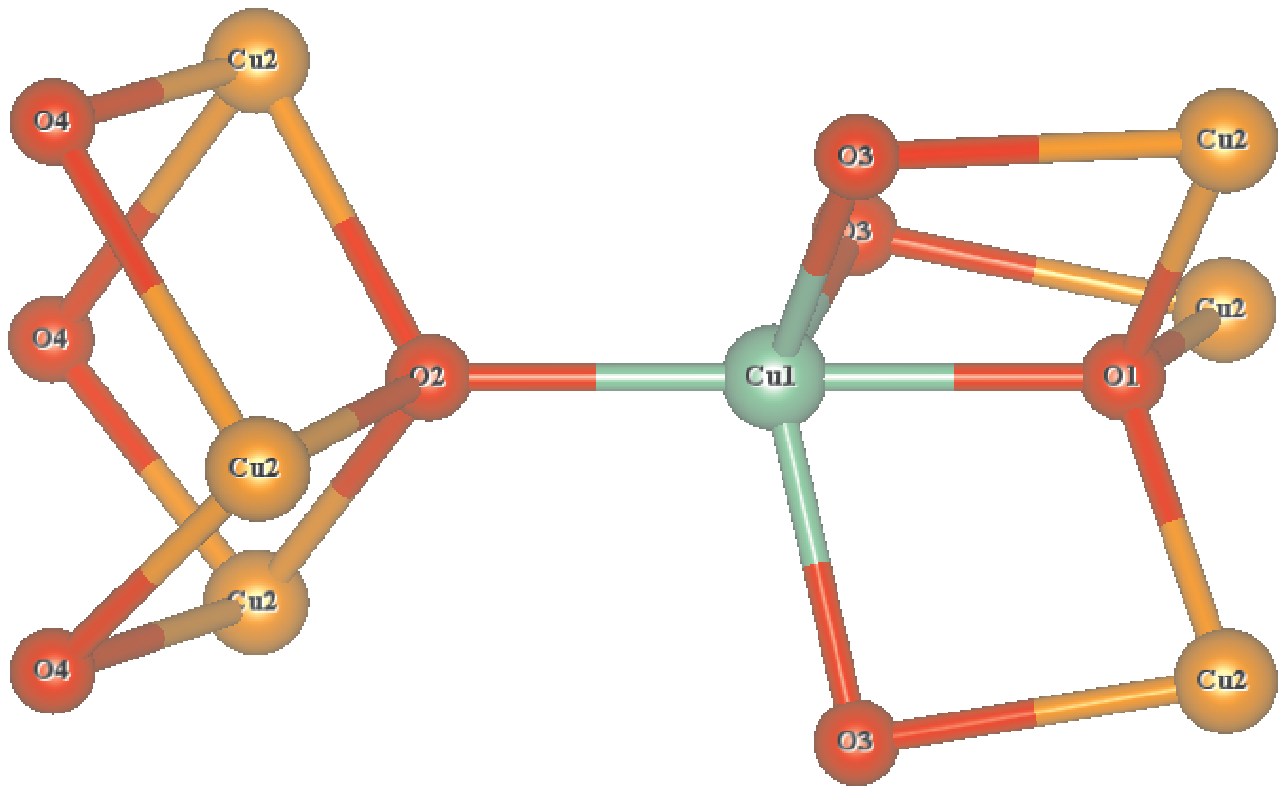}
\caption{(Color online) Local environment around $\ca$ and $\cb$ ions in $\coso$.}
\label{fig-structure}
\end{figure}

The presumed equivalence of all the AFM interactions between the central $\ca$ moment and six NN $\cb$ moments is not correct. The single bridge over the O(2) ion on one side can be estimated to be AFM with a moderate strength. However, on the other side there are two bridges, over O(1) and O(3) ions. The coupling over O(1) is also AFM but because of a smaller angle it is weaker in comparison with the O(2) bridge. Additionally, the angle over O(3) indicates the FM coupling, further reducing the total exchange constant, Fig.~\ref{fig-structure}.

Following the same arguments, the presumed single FM interaction between the NN $\cb$ moments needs to be revisited. Again there are three bridges forming two exchange interactions. For the bridge over O(1) the angle is $\sim 111^0$, indicating an AFM coupling. Since the $\cb$ moments form a triangle, there is induced frustration. A single interaction is mediated via O(2) and O(4) bridges and its sign cannot be estimated easily. Considering the angles over O(2) and O(4), the couplings should be AFM and FM, respectively, but the exact calculation is needed to determine which type is dominant.

Such a configuration of NN interactions can in principle be approximated with a mean-field theory so that on average one FM and one AFM interaction describe well the high temperature behavior. However, close to the transition one would expect that the details of the couplings start to play an important role. This can explain the overestimation of the critical temperature in the mean-field model~\cite{Belesi2010}.

Below $T_c$, in the ferrimagnetic state, the magnetization grows linearly when the dc magnetic field is applied. Our results show that all principal directions have very similar M -- H slope. We attribute this to the formation of domains along 4 equivalent space diagonals of the cube. Since each diagonal can have 2 directions of the composite moment in a ferrimagnet, we have 8 preferred domain orientations in total for $H = 0$. The application of a small magnetic field causes domain walls to move, increasing the volume of domains preferentially oriented in the direction of the field. As the temperature is lowered the pinning of domain walls becomes important, giving rise to the observed FC -- ZFC splitting. Also, depending on the orientation of the applied field, there will be one + three ([111] + [11$\bar{1}$], [1$\bar{1}$1], [1$\bar{1}\bar{1}$]), two ([111], [11$\bar{1}$]) or four ([111], [11$\bar{1}$], [1$\bar{1}$1], [1$\bar{1}\bar{1}$]) preferred domains for $H \parallel $ [111], [110] and [100], respectively. This means that there will be a value of the magnetic field $H^{meta}$ where the sample will consist of only preferred domains and for $H > H^{meta}$ the domains will start to collectively rotate towards the direction of the magnetic field. We suggest that this process is responsible for the observed metamagnetic transition in the range 200 Oe -- 500 Oe, depending on the orientation.

At a sufficiently high field a crossover to a plateau occurs, where the crossover field and the magnetization of the plateau increase as the temperature is decreased. We have found that the magnetization of the plateau is well described by a power law (see Eq.~(\ref{eq-pL})), similar to the zero-field $\mu$SR study~\cite{Maisuradze2011}. This is easily understood if we take into account that the internal fields sensed by muons are dominated by domains, irrespective of their orientation. Polarizing the sample and measuring its total magnetization should then give the same temperature dependence. However, a surprising shift of the critical temperature from 57 K in zero-field conditions~\cite{Maisuradze2011} to 60 K at $H = 1000$ Oe is observed. We can speculate that in zero field the ferromagnetic fluctuations are suppressed due to the frustration between the $\cb$ moments, producing the inflection point at 59 K and avoiding the criticality. The application of a small field is enough to disturb the delicate balance within the interactions in the system and the critical temperature rises up.

At the end we would like to discuss how our results relate to the observed magnetoelectric effect in $\coso$. The frequency-independent dielectric response suggests an absence of dielectric relaxation processes in the Hz -- MHz region, both above and below the two magnetic transitions. We were not able to observe a substantial influence of the magnetic transitions on the value of the dielectric constant. In the initial report Bos and coworkers claimed~\cite{Bos2008} that after subtraction of the lattice contribution $\epsilon_r$ exhibited a critical behavior around $T_N$. Together with the evidence that in zero magnetic field the magnetic subsystem shows avoided criticality, we find it questionable whether any kind of criticality exists in the dielectric subsystem. However, the observed maximum around 45 K and the decrease of the dielectric constant below it, which have been also observed on the powdered sample, do suggest the existence of a weak magnetoelectric effect. It is interesting to note that a sizeable magnetocapacitance of the powder occurs only below 1000 Oe~\cite{Bos2008}. We have demonstrated that this field range is characterized by the presence of domains and domain walls. Thus, we propose that the magnetoelectric effect in $\coso$ arises as a consequence of the rotation of magnetic moments within domain walls. As they spiral from one orientation to another, the (local) spatial symmetry is broken which allows the emergence of the polarization in a similar fashion as has been demonstrated in TbMnO$_3$.~\cite{Kimura2003}. Recently, it has been shown that such a scenario is realized in iron-garnet thin films~\cite{Pyatakov2011}. This way a rather small change of the dielectric constant can be naturally explained since domain walls occupy only a small fraction of the sample's volume. Further studies are needed to clarify the details of the magnetoelectric coupling.

%
%
%
%
%

\section{Conclusions}
\label{Conclusion}

We have presented a detailed single crystal magnetization and susceptibility study of $\coso$. Our results suggest that the transition from the paramagnetic to the ordered state is more complicated than previously reported. The mean-field approach, with one FM and one AFM interaction, works well at higher temperatures. However, close to the transition it is important to take into account more realistic magnetic couplings between the copper ions.

Below the transition the magnetic dynamics is dominated by the presence of domains and domain walls. A rather small field of 1000 Oe is enough to drive the system in a single-domain state, where the magnetization follows the critical behavior and a critical exponent $\beta \approx 0.39$ can be extracted, as published before~\cite{Maisuradze2011}.

The dielectric measurements showed that below the magnetic transition the dielectric constant deviates from the high temperature behavior, suggesting a weak magnetoelectric effect related to the presence of domain walls.

%
%
%
%
%

\section{Acknowledgement}
\label{Acknowledgement}

We acknowledge the financial support from Projects No.~035-0352843-2845, No.~119-1191458-1017 and No.~035-0000000-2836 of the Croatian Ministry of Science, Education and Sport and the NCCR research pool MaNEP of the
Swiss National Science Foundation.


\begin{thebibliography}{99}

\bibitem{Rado1961}
G. T. Rado and V. J. Folen.,
Phys. Rev. Lett. \textbf{7}, 310 (1961)

\bibitem{Kimura2003}
T. Kimura, T. Goto, H. Shintani, K. Ishizaka, T. Arima, and Y. Tokura,
Nature \textbf{426}, 55 (2003)

\bibitem{Hur2004}
N. Hur, S. Park, P. A. Sharma, J. S. Ahn, S. Guha, and S.-W. Cheong,
Nature \textbf{429}, 392 (2004)

\bibitem{Khomskii2009}
D. Khomskii, Physics  \textbf{2}, 20 (2009)

\bibitem{Bos2008}
J.-W. G. Bos, C. V. Colin, and T. T. M. Palstra,
Phys. Rev. B \textbf{68}, 094416 (2008)

\bibitem{Miller2010}
K. H. Miller, X. S. Xu, H. Berger, E. S. Knowles, D. J. Arenas, M. W. Meisel and D. B. Tanner,
Phys. Rev. B \textbf{82}, 144107 (2010)

\bibitem{Gnezdilov2010}
V. P. Gnezdilov, K. V. Lamonova, Yu. G. Pashkevich, P. Lemmens, H. Berger, F. Bussy, and S. L. Gnatchenko, Fiz. Niz. Temp. 36, 688 (2010) [Low Temp. Phys. 36, 550 (2010)]

\bibitem{Belesi2010}
M. Belesi, I. Rousochatzakis, H. C. Wu, H. Berger, I. V. Shvets, F. Mila and J. P. Ansermet,
Phys. Rev. B \textbf{82}, 094422 (2010)

\bibitem{Maisuradze2011}
A. Maisuradze, Z. Guguchia, B. Graneli, H. M. Ronnow, H. Berger, and H. Keller,
Phys. Rev. B \textbf{84}, 064433 (2011)

\bibitem{Koebler1999}
U. K\"obler, A. Hoser, M. Kawakami , T. Chatterji, J. Rebizant,
JMMM \textbf{205}, 343 (1999)

\bibitem{Seeger1995}
M. Seeger, S. N. Kraul, H. Kronm\"uller and R. Reisser,
Phys. Rev. B \textbf{51}, 12585 (1995)
and references therein

\bibitem{Huang2011}
C. L. Huang, K. F. Tseng, C. C. Chou, S. Mukherjee, J. L. Her, Y. H. Matsuda, K. Kindo, H. Berger and H. D. Yang,
Phys. Rev. B \textbf{83}, 052402 (2011)

\bibitem{Fujiki1981}
S. Fujiki and S. Katsura,
Prog. Theor. Phys. \textbf{65}, 1130 (1981)

\bibitem{Bitoh1993}
T. Bitoh, T. Shirane and S. Chikazawa,
J. Phys. Soc. Jpn. \textbf{62}, 2837 (1993)

\bibitem{Nair2003}
S. Nair and A. Banerjee,
Phys. Rev. B \textbf{68}, 094408 (2003)

\bibitem{Narita1996}
N. Narita and I. Yamada,
J. Phys. Soc. Jpn. \textbf{65}, 4054 (1996)

\bibitem{Bitoh1993a}
T. Bitoh, K. Ohba, M. Takamatsu, T. Shirane and S. Chikazawa,
J. Phys. Soc. Jpn. \textbf{62}, 2583 (1993)

\bibitem{Ashcroft1976}
N. W. Ashcroft and N. D. Mermin,
Solid State Physics, Saunders College, Philadelphia 1976, p 702

\bibitem{Mizuno1998}
Y. Mizuno, T. Tohyama, S. Maekawa, T. Osafune, N. Motoyama, H. Eisaki, and S. Uchida,
Phys. Rev. B \textbf{57}, 5326 (1998)

\bibitem{Pyatakov2011}
A. P. Pyatakov, D. A. Sechin, A. S. Sergeev, A. V. Nikolaev, E. P. Nikolaeva, A. S. Logginov and A. K. Zvezdin,
Eur. Phys. Lett. \textbf{93}, 17001 (2011)


\end{thebibliography}
\end{document}